\documentclass[sigconf,authorversion]{acmart}
\acmSubmissionID{1234}

\usepackage{booktabs} % For formal tables

\usepackage{graphicx}
\usepackage{booktabs} % For formal tables}
\usepackage{nicefrac}

\usepackage{tikz}
\usepackage{pgfplots}
\usetikzlibrary{shadows,backgrounds,arrows,positioning,matrix,calc}
\pgfplotsset{compat=newest}
\tikzset{
    >=stealth',
    pil/.style={
           ->,
           thick,
           shorten <=2pt,
           shorten >=2pt,}
}

\definecolor{myred}{rgb}{0.86,0.00,0.00}
\definecolor{myredlight}{rgb}{0.97,0.75,0.75}
\definecolor{myredlighter}{rgb}{0.99,0.94,0.94}
\definecolor{myredlighterr}{rgb}{1.0,0.98,0.98}
\definecolor{myblue}{rgb}{0.00,0.20,0.70}
\definecolor{mybluelight}{rgb}{0.75,0.80,0.93}
\definecolor{mybluelighter}{rgb}{0.94,0.95,0.98}
\definecolor{mybluelighterr}{rgb}{0.98,0.99,1.0}
\definecolor{mygreen}{rgb}{0.10,0.50,0.10}
\definecolor{mygreenlight}{rgb}{0.78,0.88,0.78}
\definecolor{mygreenlighter}{rgb}{0.94,0.97,0.94}
\definecolor{mygreenlighterr}{rgb}{0.99,0.99,0.99}
\definecolor{mygrey}{rgb}{0.40,0.40,0.40}
\definecolor{mygreylight}{rgb}{0.85,0.85,0.85}
\definecolor{mygreylighter}{rgb}{0.96,0.96,0.96}
\definecolor{mygreylighterr}{rgb}{0.99,0.99,0.99}
\definecolor{myorange}{rgb}{1.0,0.50,0.00}
\definecolor{myorangelight}{rgb}{1.0,0.87,0.75}
\definecolor{myorangelighter}{rgb}{1.0,0.96,0.93}
\definecolor{myorangelighterr}{rgb}{1.0,0.99,0.98}

\usepackage[ruled]{algorithm2e} % For algorithms
\usepackage{algorithmic}

\SetAlFnt{\small}
\SetAlCapFnt{\small}
\SetAlCapNameFnt{\small}
\SetAlCapHSkip{0pt}

\setcopyright{rightsretained}
\copyrightyear{2021}
\acmYear{2021}
\acmConference{SIGGRAPH '21 Talks}{August 09-13, 2021}{Virtual Event, USA}\acmBooktitle{Special Interest Group on Computer Graphics and Interactive Techniques Conference Talks (SIGGRAPH '21 Talks), August 09-13, 2021}\acmDOI{10.1145/3450623.3464645}
\acmISBN{978-1-4503-8373-8/21/08}

\begin{document}
% Title portion
\title{Lessons Learned and Improvements when Building \\ Screen-Space Samplers with Blue-Noise Error Distribution }

% DO NOT ENTER AUTHOR INFORMATION FOR ANONYMOUS TECHNICAL PAPER SUBMISSIONS TO SIGGRAPH 2019!
\author{Laurent Belcour}
\affiliation{%
  \institution{Unity Technologies}
  \country{}
  \city{}
}
\author{Eric Heitz}
\affiliation{%
 \institution{Unity Technologies}
  \country{}
  \city{}
}

\renewcommand\shortauthors{Belcour and Heitz}

\begin{abstract}
Recent work has shown that the error of Monte-Carlo rendering is visually
more acceptable when distributed as blue-noise in screen-space. Despite
recent efforts, building a screen-space sampler is still an open problem.
In this talk, we present the lessons we learned while improving our
previous screen-space sampler. Specifically: we advocate for a new criterion
to assess the quality of such samplers; we introduce a new screen-space
sampler based on rank-1 lattices; we provide a parallel optimization method that
is compatible with a GPU implementation and that achieves better quality;
we detail the pitfalls of using such samplers in renderers and how to
cope with many dimensions; and we provide empirical proofs of the versatility
of the optimization process.
\end{abstract}



% \keywords{sampling, blue-noise error distribution, rank-1 lattice}

\begin{teaserfigure}
      \tiny
      \hspace{-0.65cm}
      \begin{tikzpicture}
            \begin{scope}[shift={(-1.9cm,0)}]
                  \node at (0.95cm,3.8cm) {\small$\mathbf{s}^k_{p} = \mbox{mod}\left(\mathbf{s}^k + \mathbf{u}_{p}\right)$};
                  % \draw (0,0) rectangle (2cm, 3.5cm);
                  \begin{scope}[shift={(0,0)}]
                        \begin{axis}[height=3.2cm, width=3.2cm, xtick=\empty, ytick=\empty]
                              \addplot[only marks, color=black!60!green] coordinates {
                                    (0.605079, 0.556659)
                                    (0.105079, 0.0625)
                                    (0.855079, 0.804688)
                                    (0.355079, 0.3125)
                                    (0.730079, 0.183594)
                                    (0.230079, 0.6875)
                                    (0.980079, 0.4375)
                                    (0.480079, 0.9375)
                                    (0.667579, 0.869141)
                                    (0.167579, 0.375)
                                    (0.917579, 0.117188)
                                    (0.417579, 0.625)
                                    (0.792579, 0.492188)
                                    (0.292579, 0.0)
                                    (0.0425792, 0.75)
                                    (0.542579, 0.25)                                    
                              };
                        \end{axis}
                  \end{scope}
                  \begin{scope}[shift={(0,1.885cm)}]
                        \begin{axis}[height=3.2cm, width=3.2cm, xtick=\empty, ytick=\empty]
                              \addplot[only marks, color=red] coordinates {
                                    (0.136458, 0.556659)
                                    (0.636458, 0.0625)
                                    (0.386458, 0.804688)
                                    (0.886458, 0.3125)
                                    (0.261458, 0.183594)
                                    (0.761458, 0.6875)
                                    (0.511458, 0.4375)
                                    (0.011458, 0.9375)
                                    (0.198958, 0.869141)
                                    (0.698958, 0.375)
                                    (0.448958, 0.117188)
                                    (0.948958, 0.625)
                                    (0.323958, 0.492188)
                                    (0.823958, 0.0)
                                    (0.573958, 0.75)
                                    (0.073958, 0.25)                                                                     
                              };
                        \end{axis}
                  \end{scope}
            \end{scope}
            \begin{scope}[shift={(0,0)}]
                  \node at (1.9cm,3.75cm) {\small$\{\mathbf{u}_{p}\}_{p = (i,j)}$};
                  % \draw (0,0) rectangle (3.5cm, 3.5cm);
                  \node[rectangle, draw, inner sep=-0.0cm, anchor=south west] { \includegraphics[width=3.48cm]{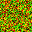} };
            \end{scope}
            \node[anchor=south] at (1.0, -0.5) {\small\textbf{a) Tile of scrambled sequences}};
            \node[anchor=south] at (1.0, -0.8) {\small\textbf{using a rank-1 lattice}};
            \begin{scope}[shift={(-0.29,2.13)}]
                  \draw[line width=0.025cm] (0, 0) -- +(0.75,0.0);
                  \draw[line width=0.025cm] (0.75,-0.05) rectangle +(0.1, 0.1);
            \end{scope}
            \begin{scope}[shift={(-0.29,1.09)}]
                  \draw[line width=0.025cm] (0, 0) -- +(0.8,0.0) -- +(0.8,0.88);
                  \draw[line width=0.025cm] (0.75,0.88) rectangle +(0.1, 0.1);
            \end{scope}
            \draw[line width=0.025cm, ->] (3.5, 2.35) -- +(1.6,0.0);
            \begin{scope}[shift={(3.5cm,0)}]
                  \node at (0.95cm,3.75cm) {\small$f(\mathbf{x})$};
                  % \draw (0,0) rectangle (2cm, 3.5cm);
                  \begin{scope}[shift={(0.5, 2.79cm)}]
                        \node[rectangle, draw, inner sep=0.0cm, anchor=south west] { \includegraphics[width=0.7cm]{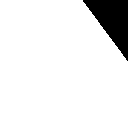} };
                  \end{scope}
                  \begin{scope}[shift={(0.5, 2.0cm)}]
                        \node[rectangle, draw, inner sep=0.0cm, anchor=south west] { \includegraphics[width=0.7cm]{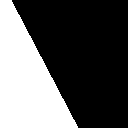} };
                  \end{scope}
                  \begin{scope}[shift={(0.85, 1.85cm)}]
                        \node {$ \vdots $};
                  \end{scope}
                  \begin{scope}[shift={(0.5, 0.8cm)}]
                        \node[rectangle, draw, inner sep=-0.0cm, anchor=south west] { \includegraphics[width=0.7cm]{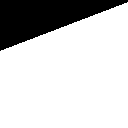} };
                  \end{scope}
                  \begin{scope}[shift={(0.5, 0.0cm)}]
                        \node[rectangle, draw, inner sep=-0.0cm, anchor=south west] { \includegraphics[width=0.7cm]{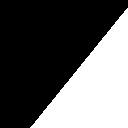} };
                  \end{scope}
            \end{scope}
            \draw[line width=0.025cm, ->] (7.8, 2.35) -- +(1.6,0.0);
            \begin{scope}[shift={(5.2cm,0)}]
                  \node at (1.65cm,3.75cm) {\small$I_N = {1\over N}\sum_k f(\mathbf{s}^k_{p})$};
                  % \draw (0,0) rectangle (3.5cm, 3.5cm);
                  \node[rectangle, draw, inner sep=-0.0cm, anchor=south west] { \includegraphics[width=3.48cm]{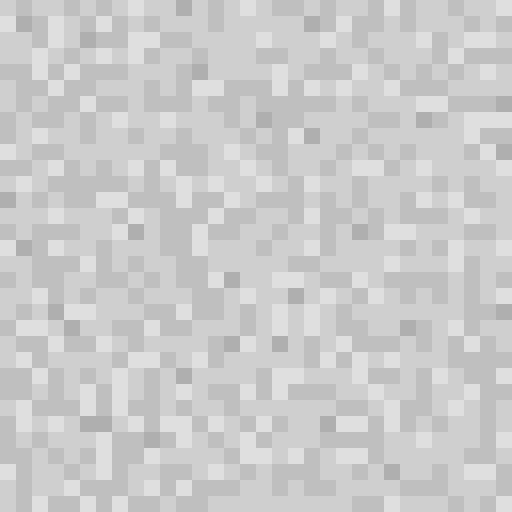} };
                  \node[rectangle, draw, inner sep=-0.0cm, anchor=south west]  at (0.1, 0.3) { \includegraphics[width=1.0cm]{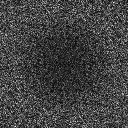} };
                  \node[inner sep=-0.0cm, anchor=south]  at (0.6, 0.1) { {FFT} };
            \end{scope}
            \node[anchor=south] at (6.9, -0.5) {\small\textbf{b) Test integrands}};
            \begin{scope}[shift={(13.2cm,0)}]
                  % \draw (0,0) rectangle (3.5cm, 3.5cm);
                  \node[rectangle, draw, inner sep=-0.0cm, anchor=south west] { \includegraphics[width=3.48cm,]{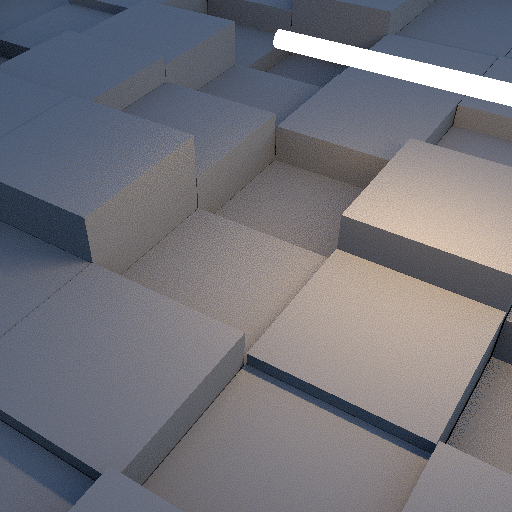} };
                  \draw (2.2,1.0) rectangle +(0.5cm, 0.5cm);
                  \node[rectangle, draw, inner sep=-0.0cm, anchor=south west]  at (0.1, 0.3) { \includegraphics[width=1.0cm]{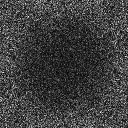} };
                  \node[inner sep=-0.01cm, anchor=south]  at (0.6, 0.1) { {\textcolor{white}{FFT of error}} };
            \end{scope}
            \node[anchor=south] at (14.8, -0.5) {\small\textbf{d) Rendering}};
            \begin{scope}[shift={(9.3cm,0)}]
                  \node at (1.65cm,3.75cm) {\small$||(I_N \circledast k_\sigma) - I_{\mbox{ref}}||$};
                  \begin{loglogaxis}[
                        height=5.09cm,
                        width=5cm,
                        grid=major,
                        xmin = 0.5, xmax = 20.0,
                        legend style={at={(1.3,0.95)}},
                        tick label style={color=white}
                  ]
                        \input{fig/teaser/whitenoise-16spp_heaviside.data}
                        \addlegendentry{Random Scrambling}
                        \input{fig/teaser/owenxor-16spp_heaviside.data}
                        \addlegendentry{\citet{heitz2019low}}
                        \input{fig/teaser/rank1cpr-16spp_heaviside.data}
                        \addlegendentry{Ours}
                  \end{loglogaxis}
            \end{scope}
            \node[anchor=south] at (10.9, -0.5) {\small\textbf{c) RMSE of test integrand}};
            \node[anchor=south] at (10.9, -0.8) {\small\textbf{w.r.t. denoising kernel size }};
      \end{tikzpicture}
      \caption{
            In this work, we optimize a tile of 2D points $\in [0,1]^2$ (a-right) to shift the value of a 2D \textit{rank-1 lattice} (a-left). We introduce a \textit{GPU-optimizer} to achieve blue-noise when integrating test integrands (b). We found that our new rank-1 screen-space sampler produces lower error than our previous sampler~\cite{heitz2019low} after \textit{denoising the test integrands tiles} (c). Finaly, we apply our sampler using \textit{dimension padding} in a path tracer to achieve blue-noise distribution of the error.
            \label{fig:teaser}
      }
\end{teaserfigure}

\maketitle

%%
%% Introduction
%%
\section{Introduction}
\label{sec:introduction}

Rendering via Monte Carlo (MC) integration is subject to numerical error that
leads to perceivable artifacts such as aliasing and/or noise in the final
image.  Traditionally, these artifacts are understood to be linked to the
\textit{amplitude} of the integration error, and can thus be attenuated via
variance-reduction techniques such as importance sampling.  Georgiev and
Fajardo~\shortcite{georgiev2016blue} showed that the \textit{screen-space
distribution} of this error also has a direct impact on the perceptual quality
of the image.

They introduced an optimization technique to build a screen-space sampler, a
sampler where each pixel of a rendered image uses a correlated sequence of
random numbers. They showed that the perceptual quality of images
improved for one-sample-per-pixel renderings. In a previous
talk~\cite{heitz2019low}, we built on this idea to support the
many-samples-per-pixel case and preserve convergence properties. We relied on
the Sobol sequence and scrambled it with XORs to generate a different sequence
per pixel.

In this talk, we continue our research effort on this topic and provide
different improvements to our previous method. First, we found it tedious to judge
rendering quality as it relies on perception and might be prone to bias. We
developped a \textit{quantifiable criterion} to assess the quality of
screen-space sampler. From this criterion, we changed \textit{the sequence and
its scrambling mechanism}. Futhermore, we improve the optimization process by
\textit{modifying the cost function} (see our supplemental material) and
\textit{parallelizing the optimizer} to benefit from GPU's performances. We
found how optimizing using discontinuous test integrands could also
\textit{benefit other classes of integrands}. Finaly, we render images using
\textit{dimension padding} for which some care is needed.

\section{Our Previous Method}
A screen-space sampler is a 2D map of quasi-random numbers we defined as
$\mathcal{S} = \left\{\mathbf{s}_{p}^k\ : p,k\right\}$, where $p$ is the pixel index $p =
(i,j)$ and $k$ is the sample index, $\mathbf{s}_{p}^k$ is a n-D vector $\in
[0,1]^n$. Following \citet{georgiev2016blue}, individual sequences of
quasi-random numbers are generated by scrambling a \textit{main sequence}
$\mathbf{s}^k$, with a \textit{scrambling function} $\mathcal{K}(\mathbf{s}^k,
\mathbf{u}_{p})$ where $\mathbf{u}_{p}$ is a random scramble value.  The
screen-space sampler is then defined by the main sequence and the 2D
\textit{tile} storing the scrambling parameters $\mathcal{U} =
\{\mathbf{u}_p\}_{p}$.  For example, in our previous implementation, we used a
Owen scrambled Sobol sequence as the \textit{main sequence} and
scrambled its values with XORs at each pixel~\cite{heitz2019low}.
To distribute the error as a
blue-noise, prior to rendering, we optimized the tile by permuting its pixels
to reduce the following loss:
\begin{align}
   \mathcal{L} = \sum_{p} \sum_{q \neq p}
                 \mbox{exp}\left( - \dfrac{\lVert(p-q\rVert)^2}{2.1^2} \right)
                 \lVert \mathcal{I}_p - \mathcal{I}_q \rVert^2,
                 \label{eq:loss}
\end{align}
where $\mathcal{I}_p = \{ \nicefrac{1}{N} \sum_k f_i(\mathbf{s}_{p}^k)\}$ is
the 2D map of \textit{test integrands} (randomly oriented Heavisides). Each pixel contains the vector of
Monte-Carlo integrals of random Heavisides functions (Fig.~\ref{fig:teaser}(b)).
Last, the order of the sequence in each pixel is also altered using a different
ranking to ensure that blue-noise is achieved for different sample counts.
In this talk we will omit this part as we did not improve it.

%%
%% Contribution Section
%%
\section{Improving a Screen-Space Sampler}

\subsection{Padding Dimensions}
The first improvement we made to the screen-space sampler was to restrict the
main sequence and scramble map to 2D values and padded this sequence to
integrate higher dimensions.  Since blue-noise is only optimized for pairs of
dimensions, dimensions must be consumed by the renderer in pairs. That is, to
integrate motion blur (1D), two dimensions must be allocated for this effect and
one of them is discarded. Because, to achieve blue-noise, neighboring pixels must use samples dimensions
for the same dimensions of the integrand. Thus, a screen-space sampler (either this
improvement or the previous one) should not be used with bidirectional path
tracing (as done by~\citet{ahmed2020screen}).

\subsection{Rank-1 Lattice}
After many tests, we opted for a different main sequence and scrambling function. We
choose the Rank-1 Lattice sequence: $\mathbf{s}^k = \mbox{mod}(\Phi(k) \times
\mathbf{d}, 1)$, where $\Phi(k)$ is the van der Corput sequence, and $\mathbf{d}$
is the \textit{direction vector} of the sequence. In our
implementation, we used the direction vectors of~\citet{hickernell2012weighted}.
A natural scrambling mechanism for such a sequence is to offset it modulo one:
$\mathcal{K}(\mathbf{s}^k, \mathbf{u}_{p}) = \mbox{mod}(\mathbf{s}^k +
\mathbf{u}_{p}, 1)$ (see Fig.~\ref{fig:teaser}(a)).  Although Sobol is
theoretically superior w.r.t. convergence, we found that rank-1 lattices
offered the same convergence for rendering integrands while
improving upon the blue-noise distribution of rendering error (see our supp. mat.).

\subsection{Evaluating a Screen-Space Sampler}
In this work, we advocate that the correct method to evaluate a screen-space
sampler is \textit{how much it reduces variance after denoising}. Indeed,
a Monte-Carlo renderer image is most of the time denoised. Denoising
consists of applying an edge-avoiding, low pass filter in screen-space. Blue-noise
error distribution does not contain energy in the low frequency
region, this permits to reduce the magnitude of the error after denoising.
A denoising kernel is driven by additional buffers (z-buffer, normal buffer, ...)
and varies in size. Thus, we apply the criterion on a range of filter sizes.
We use gaussian filters of varying standard deviation (we took $\sigma \in
[0.25, 20]$ in our examples). To ensure that our criterion is scene-independent,
we leverage again the test integrands and apply the kernel on the tile of
QMC estimates (see Figure~\ref{fig:teaser}~(c)). In this setting, we found
that our new screen-space sampler consistently performed better than our previous
XORed Owen sequence. We found the same difference when applying blurring
to rendering images (see supp. mat.). This idea bears similarity with the
work of \citet{chizhov2020perceptual}. However, they use screen-space blur
to drive the optimization, and not to evaluate the final sampler.

\subsection{Efficient Optimization}
We parallelize the optimization process by testing multiple couples of pixels
for swapping at the same time. To avoid collision, we ensure that a pixel in
a swap couple is not used in another couple. We evaluate the cost function
with no modification and perform the mutation as in the sequential optimisation.
Of course, this method can produce a tile with a
higher cost function due to concurrent swappings of neighboring pixels. However,
if we restrict the number of pixels for swapping to $\nicefrac{N}{4}$, we observe a
behavior comparable to simulated annealing: at the beginning of the optimisation, 
many concurrent swappings occur and non optimal configurations are accepted.
However, as the number of mutation increases and the number of valid swapping
candidates reduces, the algorithm accepts only a few valid mutations per cycle.
We implemented our GPU-optimizer using compute shaders. To generate the
list of swap candidates, we precompute a permutation of pixel indices that we
store in a linear array. We scramble the pixel indices using an XOR with a different 
seed per compute pass to avoid testing the same pixels again and computing another
permutation. This optimizer allowed us to generate tiles in under a minute.

\subsection{Versatility of Test Integrands}
Despite being optimized for discontinuous integrands, the screen-space sampler
achieves blue-noise distribution of integrand noise even for low frequency
integrands. We found that the vector space of test integrands shed light 
on this behavior (see our supp. mat.).

\bibliographystyle{ACM-Reference-Format}
\bibliography{siggraph}
\end{document}